\documentstyle[11pt,fullpage,setspace]{article}
\def \be{\begin{equation}}
\def \ee{\end{equation}}
\begin{document}
\onehalfspace
\begin{flushright}
TIFR/TH/95-56\\
November 1995
\end{flushright}
\bigskip
\centerline{\bf NOBEL LEPTONS \footnote{To appear in {\it
Current Science} }  }
\vskip 1.0cm
\centerline{
K. V. L. Sarma }
\centerline{ Tata Institute of Fundamental Research }
\centerline{Homi Bhabha Road, Bombay 400 005, India}
\centerline{(E-mail: \it {kvls@theory.tifr.res.in})}
\baselineskip=0.8cm

\begin{quotation} \small{The 1995 Nobel Prize in Physics is
shared equally by the American physicists Frederick J. Reines and
Martin L. Perl for their pioneering experimental contributions to
lepton physics. Following is a brief account of their discoveries.}
\end{quotation} \baselineskip=0.8cm

This is the year of the leptons. Of the six leptons of today's
particle physics, Nobel Prizes were awarded earlier for discovering
two of them: the electron and the muon-neutrino. This year it is the
turn of two more leptons, the electron-neutrino and the tau lepton, to
acquire `Nobelity'.

Frederick Reines (University of California, Irvine), aged 77, reported
the first direct evidence for the existence of the `neutrino' in 1956.
He performed the experiments jointly with Clyde L. Cowan, Jr., who
died in 1974. The neutrino is a tiny neutral particle hypothesized by
Wolfgang Pauli, Jr. way back in 1930, and was thought to be
unobservable. Twenty five years later and with the invention of the
nuclear reactor it was possible to verify the neutrino hypothesis.

Martin Perl (Stanford University, Stanford), aged 68, assisted by a
35-member team, discovered the `tau lepton' in 1975. This particle
carries a unit of electric charge and weighs approximately twice as
much as the proton. It is characterised by a shortlife of only a
fraction of a pico-second and has the distinction of being the heaviest
and the shortest-lived lepton. It is interesting that while Reines's
neutrino results were long hoped for, Perl's tau discovery came as a
complete surprise - that too at a time when people were busy
establishing the charm flavour around the same mass. The tau lepton
heralded the existence of the ``third generation'' of fundamental
constituents of matter.
\newpage
\begin{center}
{\bf {Brief History of the Neutrino}}
\end{center}
\medskip
\begin{quotation} \small{``The neutrino is the
smallest bit of material reality ever conceived of by man; the largest
is the universe. To attempt to understand something of one in terms of
the other is to attempt to span the dimension in which lie all
manifestations of natural law. Yet even now, despite our shadowy
knowledge of these limits, problems arise to try the imagination in
such an attempt.''}
\end{quotation}
\begin{flushright}
F. Reines and C. L. Cowan, Jr. \cite{RC56}.
\end{flushright}

Neutrino `was born' in a letter written by Pauli. This historic letter
dated 4 December 1930 from Z\"{u}rich, started with the unusual
greeting ``Dear Radioactive Ladies and Gentlemen'', and was read out
at the T\"{u}bingen meeting. Pauli mentioned that he had hit upon the
neutrino as a `desperate remedy' to save, among other things, the
principle of energy conservation in beta decay.  The letter referred
to the hypothetical particle as `the neutron' (the nuclear constituent
neutron was discovered later by James Chadwick in 1932). In Italian
the word `neutrone' means something like `large neutral one'. It was
Enrico Fermi who named the Pauli particle as the `neutrino', meaning
the little neutral one, and that name stuck on.

In 1931 Pauli visited Rome and discussed his proposal of the particle
with Fermi who received the idea with `a very positive attitude'.  At
the Solvay conference in October 1933, Pauli announced the neutrino to
be a particle which possessed no electric charge but carried the
missing energy and momentum and escaped the detecting equipment. The
famous beta decay theory of Fermi appeared in 1934.

It is interesting that even the great Pauli did not fully recognize
the implications of the neutrino, particularly in regard to its
penetrating power. His own account of this early period, written
sometime in 1957, is extremely fascinating and is now available in
English translation \cite{KW91}. Initially he thought he had done a
`frightful thing' as the neutrino was expected to have penetrating
power similar to, or about 10 times larger than, a gamma ray.  However
in 1934 Hans Bethe and Rudolf Peierls argued that the neutrino had to
be even more elusive as its interaction mean free path had to be
astronomical in magnitude.\footnote{For the case of reactor
antineutrinos going through water, taking the cross section to be
$\sigma (\bar {\nu _e}+p \rightarrow e^+ + n) \sim 6\times
10^{-44}~{\rm cm}^2$ and noting that there are two free proton targets
per water molecule, the interaction mean free path will be $\lambda
=(1/n\sigma ) \simeq 2.5 \times 10^{20}~$cm $\simeq 260~$light years!}

It was Bruno Pontecorvo who first suggested that the process of
inverse beta decay could be used as a way to establishing the
neutrino. In 1946 he proposed the celebrated Chlorine-Argon
radiochemical method and suggested using the source of reactor
neutrinos \cite{KW91} for the experiment. The Cl-Ar method was taken
up almost as a life-long project by the solar-neutrino pioneer Raymond
Davis, Jr. \footnote{ It is worth recalling that the Cl-Ar experiment
of Davis performed in 1956 at the Savannah River Plant reactor was a
`failure'. The result was not clearly inconsistent with
background. The null result $\sigma ({\rm reactor~} \nu +{}^{37}{\rm
Cl} \rightarrow e^-+{}^{37}{\rm Ar}) <0.9\times 10^{-45}~{\rm cm}^2~$
was interpreted to demonstrate that the reactor neutrino is not $\nu
_e$ but $\bar {\nu _e}$. Since $\bar {\nu _e}$'s are emitted in the
beta decays of neutrons and neutron-rich nuclei, a fission reactor is
an abundant source of $\bar {\nu _e}$.}

\medskip
\begin{center}
{\bf {Detection of the Reactor Neutrino}}
\end{center}
\medskip
The first neutrino reaction to be observed \footnote{ Here as is
customary the word `neutrino' is being used in its generic sense
although, strictly speaking, what Reines and his group detected were
the signals from the electron-antineutrino $\bar {\nu _e} $.} was the
inverse nuclear beta reaction
\be \bar {\nu _e}+p \rightarrow e^+ + n~ \ee driven by the
antineutrinos from the nuclear reactor. This reaction is essentially
the reverse of the neutron decay $n\rightarrow p + e^- + {\bar {\nu
_e} }~$ from which it is obtainable by transposing the $e^-$ to the
lefthand side as an antielectron, i.e., as a $e^+$, and reversing the
reaction arrow. We can deduce the energy of the incident antineutrino
from the positron momentum:
\begin{eqnarray}
 E_{\bar {\nu _e}}&=& {E_e + \Delta + [(\Delta ^2 -m_e^2)/ 2M_p]\over
1-[(E_e-p_e \cos \theta )/M_p]} \\ &\simeq & E_e+\Delta \\ &\geq &
1.804~{\rm MeV}~, \end{eqnarray} where $E_e$ is the positron energy,
$p_e$ is the magnitude of its 3-momentum, $\theta $ is the positron
emission angle with respect to the antineutrino direction, and $\Delta
= M_n-M_p= 1.293~{\rm MeV}$ is the neutron-proton mass difference (in
units $c=1$); the second step ignores neutron recoil terms of order
$(E_e/M_p)$ and the last one gives the reaction threshold.

When the antineutrino energies are in the range of a few MeV, the
positron angular distribution is more or less isotropic and the
effective cross section may be taken to be \footnote{This implies an
approximate quadratic dependence on the incident neutrino energy
$E_{\nu }$; at higher energies (above GeV) the cross section will be
independent of $E_{\nu }$ due to the rapid decrease of the vector and
axial-vector form factors. However the total cross section increases
linearly with $E_{\nu }$ , as a consequence of the point nature of the
target quark. At still higher energies (where the $W$ mass effects
become important) the total cross section is expected to increase
logarithmically, as is indicated in the recent HERA experiments $ep
\rightarrow \nu_e +~...~$.  }
 \be \sigma = {4{G_F^2}\over \pi }p_e E_e~, \ee where $G_F $ is the
Fermi coupling constant; for $E_{\bar {\nu _e}}=$ 2.3 MeV the
cross section is
 \be \sigma \simeq 6 \times 10^{-44}~{\rm cm}^2~.\ee

The idea underlying the Nobel-Prize winning experiment was to look for
a pair of scintillator pulses, the first (prompt) pulse due to
positron annihilation and the second (delayed) one due to capture of
the moderated neutron. The experiment \cite{CO56} was performed around
1955-56. Projectiles were the reactor neutrinos from the Savannah
River Plant located in South Carolina State, USA, and targets were the
protons in a solution of water mixed with cadmium chloride (Cd is a
good absorber of thermal neutrons). The experimental apparatus could
be viewed as a multiple-sandwich with 2 layers of cheese (target
material) arranged between 3 pieces of bread: the cheese layers were
the two target tanks containing a solution of water and CdCl$_2$,
while the bread pieces were the three liquid scintillator detectors.

An event meant the detection of two prompt coincidences (see Figure
1): the first one was between the two photons (each having 0.511 MeV
energy) of the positron annihilation, and the second prompt
coincidence was due to the capture of neutron by cadmium giving a few
photons (with total energy of 9 MeV). The second pulse occurred after
several micro-seconds of the positron flash, the time it took for the
neutron to be thermalised in the target water. The experiment
involved, among other things, measuring the energies of the pulses,
their time-delays, dependence of event rate with neutrino flux which
varied with the reactor power output, etc.

The results were published in the 5-author paper C.L. Cowan, Jr., {\it
et al} \cite{CO56} entitled ``Detection of the Free Neutrino: A
Confirmation''. This paper, unusual for the lack of diagrams of the
detector or the customary figures or tables, is regarded as the
discovery paper \footnote{The earlier 1953 experiment of Reines and
Cowan [Reines, F. and Cowan, Jr. C.L., {\it Phys. Rev.}, 1953, {\bf
90}, 492; {\bf 92}, 830] performed at the Hanford reactor in
Washington, did not give a signal well above the background that was
unrelated to the reactor. The number of delayed coincidences had a
large uncertainty, 0.41$\pm$ 0.20 per hour.}. Photographs of the
detector assembly and associated equipment can be seen in
Ref.\cite{RC57}. The observed signal varied with the reactor-power. It
consisted of an average rate of $2.88 \pm 0.22~$ counts/hour,
consistent with a value $\sim 6 \times 10^{-44}~ {\rm cm}^2$ for the
inverse beta-reaction cross section. The signal to reactor-unrelated
background ratio was 3 to 1.

It is remarkable that in the same year Reines and Cowan \cite{RC56}
also gave an upper limit on the neutrino magnetic moment, \[ \mu
_{\bar {\nu _e}}< 10^{-9} \mu _B~,\] where $\mu _B$ is the Bohr
magneton. This limit was deduced from the extent of non-observation of
scintillator pulses along the path of the reactor neutrino, and is
about the same as the present upper limit which is extracted by using
different considerations (for a survey of the recent results in
neutrino physics, see e.g., Ref.\cite{KS95}).

Even today, we do not know very much about the electron-neutrino (and
much less about other neutrinos). As for its rest mass, data on the
end-point of the beta electron spectra only show that it does not
exceed a few eV; a recent high statistics experiment \cite{BE95} using
molecular tritium in gaseous form, gives the upper limit \be m(\bar
{\nu _e}) < 4.35~{\rm eV}, \ee at 95\% confidence level.

An important feature of $\nu _e$ undoubtedly is that it is
`left-handed': The spin of the neutrino, which has a magnitude of half
a natural unit, is observed to be aligned {\it opposite} to the
momentum direction. This inherent property of the neutrino was
discovered originally in the exceedingly clever experiment performed
in 1958 by Goldhaber {\it et al} \cite{GGS}. The neutrino that was
emitted in the beta decay of $^{152}$Eu by electron capture was found
to be lefthanded. This conclusion was arrived at as the photon emitted
by the excited daughter was observed to be left-circularly polarized.

Experimental data using the $\nu _e$'s are not extensive because it is
difficult to obtain $\nu _e$ beams. At the pi-meson factories, low
energy $\nu _e$ get produced in 3-body decays and that too admixed
with the muon-neutrinos ($\nu _{\mu }$ and $\bar \nu _{\mu }$). Sun is
no doubt a good source of $\nu _e$ but solar neutrino fluxes are not
well understood and constitute one of the challenging problems of
current research. Only recently the first experiment using a man-made
source of pure $\nu _e$ was performed. It made use of an intense
source of reactor-produced $^{51}$Cr (which emits $\nu _e$ by electron
capture) to calibrate the GALLEX solar neutrino detector \cite{AN95}.

\bigskip
\begin{center}
{\bf {Discovery of the Tau Lepton} }
\end{center}
\medskip

The tau lepton broke on the scene unexpectedly. While the results of
Reines needed the construction of power reactors, the discovery of tau
lepton needed high energy electron-positron colliders. Tau lepton is
the third kind of charged lepton that exists in Nature, the other two
being the electron and the muon. (The Greek letter $\tau$ is the first
in the word triton, meaning third). Its birth can be traced to the
1975 paper entitled, `` Evidence for Anomalous Lepton Production in
$e^+-e^-$ Annihilation'' by Perl {\it et al} \cite{MP75}.

The experiment was performed at the electron-positron collider called
SPEAR (Stanford Positron Electron Accelerator Ring). At this facility
beams of $e^- $ and $e^+$ were accelerated simultaneously in opposite
directions in a ring and made to intersect. As the total
center-of-mass energy (the sum of beam energies) \begin{eqnarray}
E_{cm}&=& E_{e^-}+E_{e^+}\\ &=& 2E_{e^-}~,\end{eqnarray} was tunable
in the range 3-8 GeV, a pair of charged particles each having a rest
mass of about 2 GeV could easily be produced at SPEAR.

A large cylindrical detector placed in a magnetic field surrounded the
collision area. Electrons were identified by the electromagnetic
shower counters, and muons by their ability to penetrate large amounts
of iron and other materials making up a total of 1.7 absorption
lengths for pions. Particles emitted at the polar angles between
$50^0$ and $130^0$ and at all the azimuthal angles, were recorded.

{\it Anomalous Events}: The `anomalous' events reported by Perl and
collaborators corresponded to the following reactions which had a very
distinctive signature,
\begin{eqnarray} e^-+e^+ & \rightarrow & e^- + \mu ^+ +~ i.p.~~,\\
        & \rightarrow & e^+ + \mu ^- +~ i.p.~~, \end{eqnarray} where
`$i.p.$' denotes invisible particles which left no trace in the
detector. The ingenuity of the experimenters consisted in establishing
that the oppositely-charged $e\mu $ pair was the result of separate
decays of two new particles which were oppositely-charged and
short-lived. In an effort to reduce the background due to the copious
production of $e^+e^-$ and $\mu ^+\mu ^-$ pairs, the events were
chosen to be `acoplanar' so that production of more than two particles
was ensured. For this the $\mu $ was required to make an angle more
than $20^0$ to the plane that contained the final $e$ and the incident
beam.

The very first report of Perl {\it et al} \cite{MP75} had a total of
64 anomalous events at the SPEAR range of energies. For example, at
the energy $E_{cm}=$ 4.8 GeV there were 24 events (13 $e^+\mu ^-$ and
11 $e^-\mu ^+$) with an estimated background consisting of about 5
events (arising from possible misidentification of hadrons as leptons,
decays of known hadrons into leptons, etc). The events were found to
be `noncollinear', meaning that the angle between the $e $ and $\mu $
momenta was more than $90^0$ and the two particles were emitted in
opposite hemispheres with respect to the beam.

{\it {Threshold Behaviour}}: The occurrence of anomalous events as a
function of $E_{cm}$ exhibited an increase around 4 GeV, as shown in
Figure 2. This indicated the existence of a threshold for the
production of anomalous events. Moreover, in producing a pair of
point-particles $\tau ^+ \tau ^-$ by one-photon exchange \be
e^++e^-\rightarrow \gamma _{\rm virtual}\rightarrow \tau ^+ +\tau ^-
{}~,\ee quantum electrodynamics tells us that the total cross section
$\sigma _{\tau \tau }$ should depended on the final velocity $\beta $
as follows:
\begin{eqnarray}
\sigma _{\tau \tau }& =& {(3-\beta ^2) \over 2}\beta \sigma _0~~
 ({\rm for~spin~}{\frac{1}{2}}),\\
                    & =& {\beta ^3 \over 4}\sigma _0~~~~~~~~~~~
 ({\rm for~spin~}0), \end{eqnarray}
where
\be  \beta =
\sqrt{1- {4 M_{\tau }^2 \over E_{cm}^2} }~;~ \sigma _0 \equiv {4\pi
\alpha ^2 \over 3E_{cm}^2}~;~ \alpha \simeq {1\over 137}~.  \ee
The $\beta ^3$ dependence in the boson case arises from the $p$-wave
production which is a result of parity conservation in electromagnetic
interactions. Subsequent experimental observations were consistent
with a linear rise of cross section with the velocity, and thus for
the $\tau ^{\pm }$ the assignment of spin $\frac{1}{2} $ was preferred
over spin $0$ and other possibilities.

Final interpretation of the anomalous events followed soon
\cite{MP76}: electron-positron annihilation gives rise to a pair of
tau leptons \be e^+ + e^- \rightarrow \tau ^- + \tau ^+ ~\ee which
decay immediately into lighter leptons. The `tau-lepton number'
conservation is respected by assuming the emission of an associated
neutrino called the tau-neutrino ($\nu _{\tau }$) in the $\tau ^-$
decay, and a tau-antineutrino ($\bar {\nu _{\tau }}$) in the $\tau ^+$
decay. Thus to explain the anomalous event with, say, $e^-\mu ^+$ we
appeal to the decays \be \tau ^- \rightarrow e^-+\bar {\nu _e}+ {\nu
_{\tau } }~,\ee \be \tau ^+ \rightarrow \mu ^++\nu _{\mu }+ {\bar {\nu
_{\tau }} }~,\ee wherein the two neutrinos and the two antineutrinos
(belonging to all three neutrino generations) constitute the invisible
particles.

Data gathered over the years have shown that the {\it shape} of the
electron (and also muon) energy spectrum in tau decay is in good
agreement \cite{MP77} with the standard $V$-$A$ theory. Also all the
observations support the view that the earlier $e$-$\mu $ universality
is extendible to $e$-$\mu $-$\tau $ universality. Thus the original
enigma regarding the existence of muon is deepened. It now becomes the
so-called `generation puzzle': among the ultimate constituents of
matter, why do members of one generation behave exactly in the same
way, except for the mass, as the corresponding members of another
generation?

It may be interesting to recall that in 1974 the spectacular discovery
of $J/\psi $(3.097 GeV) immediately brought forth in its wake an
intense frenzy of activity relating to the charm flavour. People were
busy studying the spectroscopy of hidden-charm states and open-charm
states and their decays. Consequently the feeling prevalent at the
time was that Perl's anomalous events were some obscure manifestation
of charm decays and the situation would be clearing up soon. However
such feelings were firmly dispelled by 1977 when the Double Arm
Spectrometer (DASP) group \cite{BR78} working at DESY, Germany,
reported seeing the anomalous $e\mu $ events around $\psi '$(3.686
GeV) manifestly below the open-charm threshold (2$\times $1.87 GeV).

The present best value for the tau lepton mass comes from the
measurements at the Beijing Electron Positron Collider (BEPC)
\cite{BA95}, and is given by \be m_{\tau }=
1776.96^{~+0.18~+0.25}_{~-0.21~-0.17}~{\rm MeV},\ee where the first
error is statistical and the second is systematic. The relatively
large mass of the tau implies a large phase space for decay, and this
makes a variety of final states to be accessible for the tau to
decay. Some of the decay states contain 5 charged particles (pions or
kaons) and possibly one or two neutral pions, besides the missing
tau-neutrino. The availability of several competing channels for decay
(e.g., the probability for the 2-body decay $ \tau ^- \rightarrow \pi
^-+\nu _{\tau }$ is $(11.7 \pm 0.4)$\%), makes the tau a very
shortlived particle. Its meanlife is presently known to 1\% accuracy
\cite{PDG}, \be \tau _{\tau } = (2.956\pm 0.031)\times 10^{-13}~{\rm
s}~.\ee

Finally, in regard to the mass of the tau-neutrino the information
available is very limited: the upper limit obtained from a recent
study \cite{BU95} of the tau-decay events, each containing 5 charged
pions, is $m(\nu _{\tau } ) < 24$ MeV at 95\% confidence level.
\bigskip
\begin{center}
{\bf {Summary}}
\end{center}
\medskip
The announcement of this year's Nobel award in Physics is in
recognition of two land-mark experiments in elementary particle
physics. One provided the first confirmation of the neutrino, as
envisaged by Pauli more than two decades earlier, as a tiny neutral
particle emitted in beta decay. The experiment made an ingenious use
of time-correlations of scintillator pulses to ascertain the occurrence
of the inverse scattering reaction. The other experimental
investigation which shared this year's honour is the discovery of the
heavy charged lepton $\tau $. This was a serendipitous discovery
perhaps like that of the muon. The `anomalous' events with the
oppositely-charged $e$ and $\mu $ were interpreted as resulting from
the independent decays of the shortlived pair $\tau ^+\tau^-$ which
was created in the reaction $e^++e^-\rightarrow \tau ^++\tau^-$. The
$\tau $ is a member of the third (and perhaps the last) generation of
ultimate constituents of matter to which the top and bottom quarks
belong (see, e.g., \cite{KVLS}).
\newpage
\centerline{ {\bf Table 1}: Lepton Generations }
\medskip
\begin{tabular}{llll}
\hline
\hline
Generation & ~~Lepton &      Discoverer(s) &   Nobel Prize Winner(s)\\
           & ~~       &      (year) &          (year) \\
\hline
&&&\\
 ~~~1&        electron $e$ &  J.J. Thomson &         J.J. Thomson  \\
   &                       &  (1897)       &         (1906)  \\
  & electron-neutrino $\nu _e$ & C.L. Cowan {\it et al} \cite{CO56} & F.J.
Reines\\
  &                       &     (1956)   &             (1995) \\
   &&&\\
 ~~~2&  muon $\mu $ &   J.C. Street, E.C. Stevenson;&          \\
   &              &   C.D. Anderson, S.H. Neddermeyer & ~~ ---------- \\
   &              &       (1936)                     &        \\
   & muon-neutrino $\nu _{\mu }$ & G. Danby {\it et al} \cite{GD62}& M.
Schwartz,\\
   &     &  (1962)                    &   L.M. Lederman,\\
   &                     &                            &  J. Steinberger\\
   &                     &                            &    (1988) \\
   &&&\\
 ~~~3& tau $\tau $ &  M.L. Perl {\it et al} \cite{MP75} &  M.L. Perl \\
   &                     &  (1975)                      &    (1995)  \\
   & tau-neutrino $\nu _{\tau }$&~~~?                   &       ~~~?  \\
   & & & \\
\hline
\hline
\end{tabular}
\bigskip

A summary of the discoveries made in the world of leptons is given in
Table 1. We see that the third generation has started getting Nobel
prizes. It is amusing that the charged-leptons crop up with a 39-year
gap and may be the 4th one would show up in the year 2114. For the
present, the available experimental information implies that there are
no charged leptons which are heavier than tau and lighter than 45 GeV.

Finally it should be emphasised that the third kind of neutrino $\nu
_{\tau}$ still needs to be identified experimentally (and hence the
question marks in the last row of the Table). To this end one ought to
demonstrate that $\tau $'s are produced directly in collisions of $\nu
_{\tau}$ with nuclear targets. However, $\nu _{\tau}$ beams are hard
to obtain mainly due to the very short intrinsic lifetime of the $\tau
$.

\centerline{\bf {FIGURE CAPTIONS}}

Figure 1: Schematic diagram of the neutrino detector of the
Reines-Cowan experiment, Ref. \cite{RC57}.

Figure 2: The {\it observed} cross section for the anomalous $e-\mu $
events found by Perl {\it et al}, Ref. \cite{MP75}.
\end{document}